\documentclass[twocolumn,english,notitlepage,superscriptaddress,nofootinbib]{revtex4-1}
\usepackage[T1]{fontenc}
\usepackage[latin9]{inputenc}
\usepackage{float}
\usepackage{amsmath}
\usepackage{amssymb}
\usepackage{graphicx}
\usepackage{wasysym}
 \usepackage{color}
\makeatletter

\begin{document}

\title{Hyperpolarisation of nuclear spins: polarisation blockade }

\author{O.T. Whaites} \affiliation{Department of Physics and Astronomy, University College London,
Gower Street, London WC1E 6BT, United Kingdom}

\author{C.I. Ioannou} \affiliation{QuTech, Delft University of Technology, PO Box 5046, 2600 GA Delft, The Netherlands}
\affiliation{Kavli Institute of Nanoscience Delft, Delft University of Technology, PO Box 5046, 2600 GA Delft, The Netherlands}
\author{B.J.Pingault} \affiliation{QuTech, Delft University of Technology, PO Box 5046, 2600 GA Delft, The Netherlands}
\affiliation{Kavli Institute of Nanoscience Delft, Delft University of Technology, PO Box 5046, 2600 GA Delft, The Netherlands}
\author{G.L. van de Stolpe} \affiliation{QuTech, Delft University of Technology, PO Box 5046, 2600 GA Delft, The Netherlands}
\affiliation{Kavli Institute of Nanoscience Delft, Delft University of Technology, PO Box 5046, 2600 GA Delft, The Netherlands}
\author{T.H. Taminiau} \affiliation{QuTech, Delft University of Technology, PO Box 5046, 2600 GA Delft, The Netherlands}
\affiliation{Kavli Institute of Nanoscience Delft, Delft University of Technology, PO Box 5046, 2600 GA Delft, The Netherlands}

\author{T.S. Monteiro} \affiliation{Department of Physics and Astronomy, University College London,
Gower Street, London WC1E 6BT, United Kingdom}

\begin{abstract} 
Efficient hyperpolarisation of nuclear spins via optically active defect centers,  such as the nitrogen vacancy (NV) center in diamond,  has great potential for enhancing NMR
based quantum information processing and nanoscale magnetic resonance imaging.  Recently,  pulse-based protocols have been shown to  efficiently transfer optically
induced  polarisation of the electron defect spin to surrounding nuclear spins- at particular resonant pulse intervals.  In this work,  we investigate the performance of these protocols,  both analytically and experimentally,  with the electronic spin of a single NV defect.   We find that whenever polarisation resonances of  nuclear spins  are near-degenerate with a `blocking' spin,  which is single  spin with stronger off-diagonal coupling to the electronic central spin,    they are displaced out of the central resonant region-  without,  in general,  significant weakening of the resonance.   We analyse the underlying physical mechanism  and obtain a closed form expression for the displacement.  We propose that spin blocking represents a common but overlooked  effect in hyperpolarisation of  nuclear spins and suggest solutions for improved protocol performance in the presence of (naturally occurring) blocking nuclear spins.

\end{abstract}

\maketitle

\section{Introduction}
There is significant current interest in techniques for the control of nuclear spins using solid-state defects like nitrogen vacancy (NV) centers in diamond \cite{Degen2014,Renbao2014}.  Many of these techniques rely on protocols of periodically applied microwave pulses.  Although they were originally developed to dynamically decouple the electron spin from the environment \cite{BarGill,Zhao2011,Zhao2012a},   it was subsequently found that when pulses are applied at intervals resonant with surrounding $^{13}$C precession frequencies, the resulting entanglement between and individual nuclear spin and the electronic spin of the defect offers a very effective technique for sensing and controlling nuclear spin states \cite{Degen2014,NV3,Neumann2010}.

Such pulse-based control has been exploited for nuclear polarisation and state initialisation with 
applications ranging from quantum error correction and quantum information \cite{NV3, Neumann2010,Bernien2013, Taminiau2014,Retzker,Registers},  to nanoscale NMR and other sensing applications
\cite{Wrachtrup2017,Ajoy2022}.  Dynamical nuclear polarisation (DNP) \cite{Abragam1978},   the transfer of polarisation from electrons to nuclear spins,  originally developed for NMR is also being developed in this context.  Recently proposed pulse-based DNP protocols explicitly aimed at nuclear polarisation with NVs,   PulsePol \cite{PulsePol,PulsePol1} and PolCPMG  \cite{PolCPMG} have been demonstrated to polarise $^{13}$C nuclei in diamond.  Polarisation of spins external to the diamond sample using PulsePol has been achieved  using an ensemble of NV centers \cite{Healey2021}.

\begin{figure}[ht!] 
\includegraphics[width=3.5in]{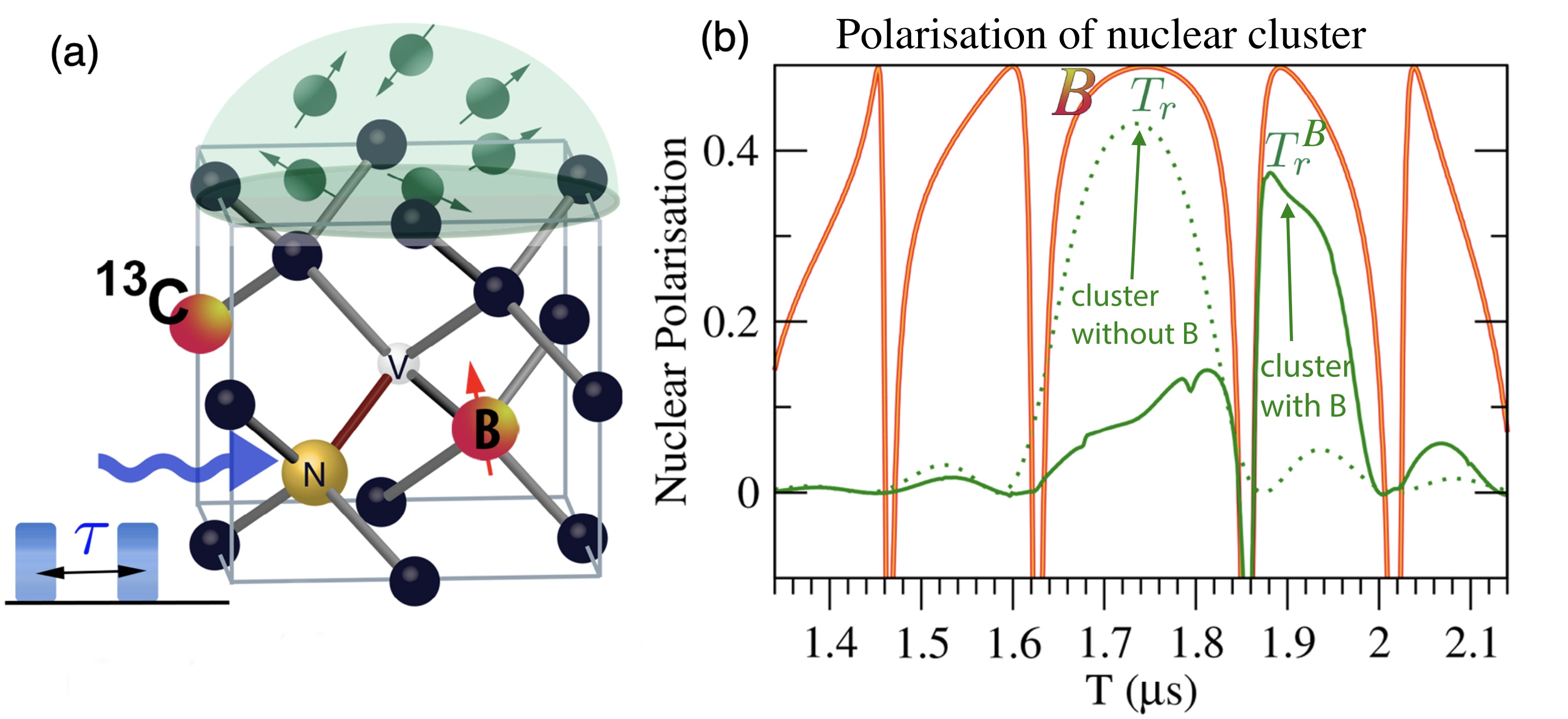}
\caption{Illustration of nuclear spin polarisation in the presence of a blocking spin: {\bf (a)} shows an NV center in diamond.  A pulse-based polarisation protocol, characterised by a pulse interval $\tau$,   is applied to polarise a distant cluster of nuclear spins (in light green) {\bf (b)} Efficient polarisation of the weakly-coupled spins (green dots) is expected near the resonant pulse period $T = T_r$.  However,  in the presence of a spin that is near-degenerate but interacts more strongly with the NV  (blocking spin B),  the cluster's resonances are displaced,  to $T \simeq T^B_r$  (solid green line).  The resonance of  spin B (red line) is unperturbed.  We term this effect `polarisation blockade,'  in analogy to blockade effects encountered in other fields of physics \cite{blockade2009}.}
\label{Fig1}
 \end{figure}
 
 However,  while these protocols were designed in the setting of polarisation transfer to a single spin,  in a realistic setting,  the central spin  couples to multiple spins.  In this study,  we investigate polarization transfer from a single spin simultaneously coupled to several environmental spins.  In particular, we find that polarisation transfer,  at the expected  polarisation resonance $T=T_r$,   can be suppressed by a blocking spin,  which is a simultaneously coupled environmental spin,  with similar precession frequencies but  with stronger coupling.  This  effect is illustrated in Fig.\ref{Fig1}: the presence of the blocking spin expels
  the polarisation resonances of the weaker-coupled spins from  the central $T\approx T_r$ region.   We  analyse the underlying physical mechanism and clarify its relation to dark states,  which are  known to suppress polarisation of nuclear spins \cite{TimeCrystal2021,DarkState}.  Spin blocking is quite distinct and,  to our knowledge,  not previously investigated: while dark-states suppress polarisation by decoupling a subspace of states from the dynamics,  spin-blocking acts by shifting a subset of spins off-resonance.   We experimentally verify this analysis using a single NV whose microscopic spin environment has been precisely characterised \cite{Abobeih2019,vandS2023}.
  
In section II we review pulse-based control of nuclear spins and  introduce the theoretical Floquet-based models used to analyse the joint dynamics of the NV and nuclear spins and simulate the experiments.  In section III we present our results,  including theory-experimental comparisons as well as an expression for the resonance displacement,  equation \ref{Displ},  a key new result of the work.  In Sec.IV we discuss the implications for nuclear polarisation and more efficient control of nuclear registers. 

\section{Methods: Pulse-based control}
An NV electron spin system surrounded by $N_{\mathrm{nuc}}$ nuclear spins may be described by the Hamiltonian ${\hat H}(t) = {\hat H}_p(t)+{\hat H}_0$ where :
\begin{equation}
{\hat H}_0 =   \omega_\text{L} \sum^{N_{\mathrm{nuc}}}_{n=1}  {\hat I}^{(n)}_z + {\hat S}_z\sum^{N_{\mathrm{nuc}}}_{n=1} \textbf{A}^{(n)} \cdot  {\bf I}^{(n)}. 
\label{Ham}
\end{equation}
The operators for the electronic spin in subspace $\{|0\rangle,|-1\rangle\}$ and the nuclear spins are labelled $\hat{S}$, $\hat{I}$ respectively.  $\omega_\text{L}$ is the nuclear Larmor frequency; the hyperfine field  $\textbf{A}^{(n)}$ acting on the nuclear spin has components $A^{(n)}_\perp, A^{(n)}_z$ relative to the $z$-axis; we take ${A^{(n)}_\perp} \equiv A^{(n)}_x$,  without loss of generality.   ${\hat H}_p(t) = \Omega(t){\hat S}_k$ is the pulse control Hamiltonian.  $\Omega(t)$ is set by the microwave control field,   while $k\equiv x,y$ for common protocols.
 
 For pulse-based control,  there is a resonant pulse spacing $\tau=\tau_r $ for which the electron and nuclear spin states selectively interact,  allowing efficient control of the nuclear states.  For example,  in the well-known CPMG sequence \cite{Schweiger},     microwave pulses are applied along the $x$-axis at regular intervals,  $\tau$; its resonant pulse spacing  is $\tau_r =j \pi/\omega_I$, where $j$ is an odd integer and the resonant frequency of a given nucleus is $\omega_I\simeq\omega_L-A_z/2$.  For common protocols, the  full protocol period $T$  is an integer multiple of $\tau$.  For CPMG specifically,  it  is $T_r= 2\tau_r$.  Note that we omit the $n$ superscript for single nuclear spin calculations.
 
\begin{figure*}[ht!]
\centering
\includegraphics[width=5.8in]{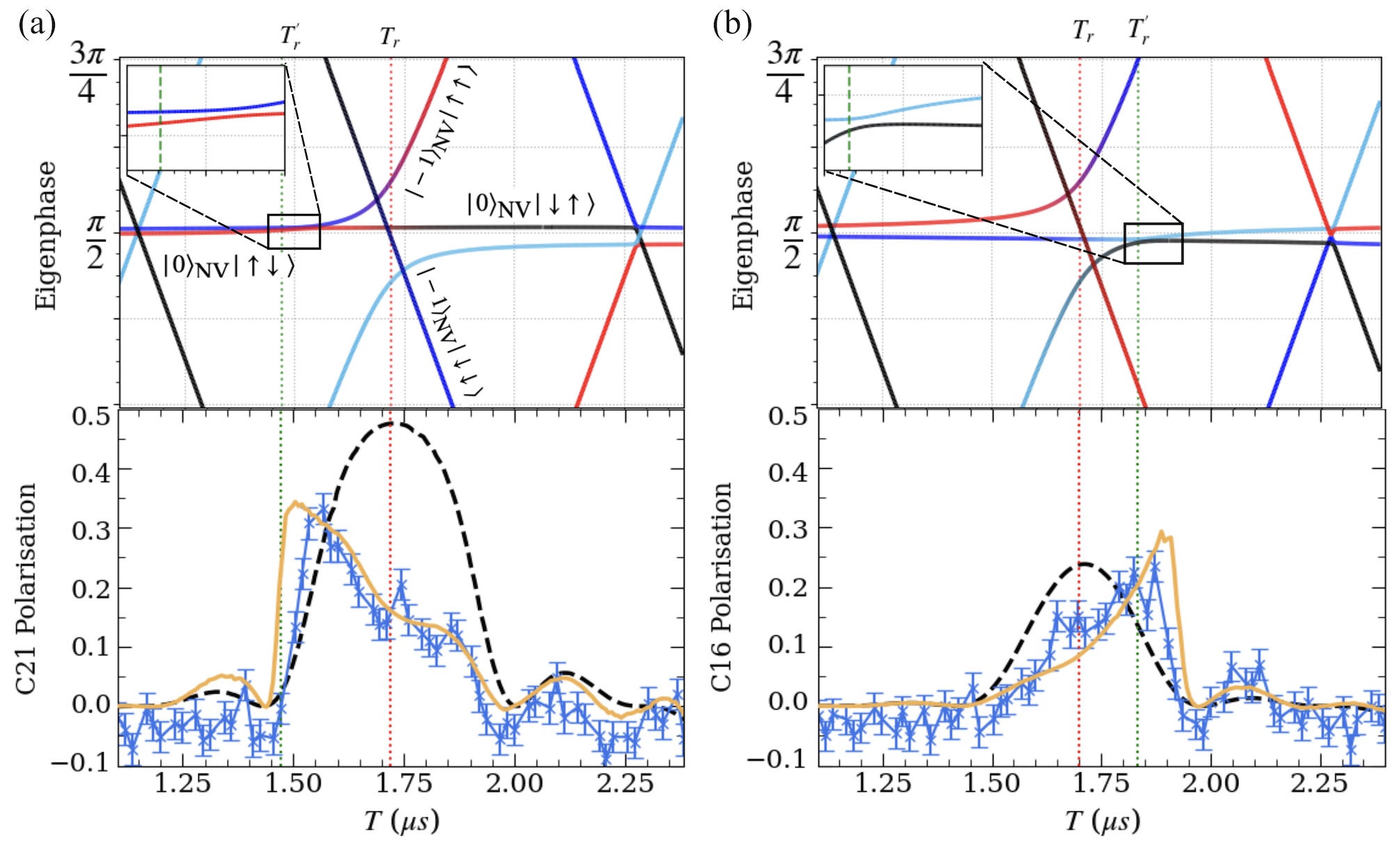}
\caption{  Comparisons with experiment and spectral analysis  {\bf (a)} Polarisation of  weakly coupled nuclear spin C21 with $(A_x,A_z)\equiv (\approx 5.0,-9.7)$ kHz $\times 2\pi$ employing the PulsePol protocol,  with 
 $N_p=4,  R=100$.  The NV environment contains a blocking spin C3 with
  $(A^B_x,A^B_z)\equiv (59.,-11.3)$ kHz $\times 2\pi$.   Nuclear spin C21 shows the  asymmetric displaced resonance expected for $N_p=4$.  Lower panels show experiment (blue),  single-spin simulation of C21 (black dash line),  simulation of C21 and C3 (orange line). The upper panel shows the corresponding Floquet spectra,  and offers an intuitive spectroscopic understanding of  spin blocking.  The wide,  broad avoided crossing is associated primarily with C3.  It overlaps with the much narrower avoided crossing corresponding to the polarisation resonance of C21.  This means the narrow C21 crossing is pushed away from $T_r$ to {\em lower}  $T'_r <T_r$.  {\bf (b)} For  C16,   with $(A_x,A_z)\equiv (5.3, -19.8)$ kHz $\times 2\pi$,   the overlap with the strong C3 avoided crossing results in the narrow C16 crossing being pushed towards larger $T'_r >T_r$.   Eq.\ref{Displ} quantifies the magnitude and clarifies that the sign of  the displacement depends on $A^B_z-A_z$.   }
 \label{Fig2}
\end{figure*}

{\em Polarisation protocols:}  recently,  new  pulse-based protocols  were identified \cite{PulsePol,PolCPMG} that split the resonance,  such that each component selectively addresses one nuclear spin state,   allowing polarisation.  In the present work we focus on the DNP protocol PulsePol,  commonly used due to its robustness to detuning \cite{PulsePol}.   PulsePol combines a series of $x$- and $y$-directional MW pulses to map the NV state onto a nuclear spin.  It has pulse period $T=4\tau$ where $\tau$ is the pulse interval.   Its  resonant pulse interval is $ \tau_r = j\pi/(2\omega_L)$, for $j=1,2...$ and the  3$^\mathrm{rd}$ harmonic ($j=3$)  is often selected for its effectiveness \cite{PulsePol}.   By averaging over the period,  an effective,  time-independent single nuclear-spin Hamiltonian  
${\hat H}^{(n)} \equiv  g^{(n)} ( {\hat S}_+{\hat I}^{(n)}_- + {\hat S}_-{\hat I}^{(n)}_+)$ can be obtained,  corresponding to a flip-flop type interaction between the NV spin ($S$) and the $n$-th nuclear spin ($I$).   The corresponding flip-flop rate at this resonance was found to be  $g^{(n)} =A_x^{(n)} (\sqrt{2}+2)/(6\pi)$, \cite{PulsePol}. 

{\em Repetitions:}  in general,   in order to achieve high levels of  polarisation,  repetitions of the polarisation protocol are required.  For each repetition,  $N_p$ cycles are applied at each value of $\tau$.  This is the periodic component,  where the NV-nuclear evolution is largely coherent.  However,  the NV electronic spin is reinitialised optically after $N_p$ cycles to its $m=0$ state.  The reduced nuclear bath evolution is (ideally) uninterrupted.  The $N_p$ sequence is then repeated. 

A series of NV reinitialisations,  interspersed with $N_p$ protocol cycles,  is repeated $R$ times.
 Typically,  protocols employ a short run $N_p=2-8$ that  yields appreciable polarisation for  only the most strongly coupled spins;  but if this is  followed by many $R \gg 1$ repetitions,  polarisation of  even nuclear spins with weak coupling is gradually achieved.   Corresponding theoretical simulation involves $R$ sets of coherent Hamiltonian evolution for $t=N_pT$  interspersed with  calculation steps where the NV states are traced out  in order to simulate re-initialisation in $m_s=0$.
\subsection{Experimental set-up}

We study nuclear polarisation dynamics surrounding a single NV center at cryogenic temperatures ($4\,\mathrm{K}$). The NV electron spin is initialised and read out via resonant optical excitation. 
The NV sample employed here was previously characterised in detail,  allowing for accurate modelling of the microscopic nuclear environment \cite{Abobeih2019, vandS2023}. 
The individual nuclear spins are labelled as C1, C2...Cn and their hyperfine coupling strengths  $A^{(n)}_x,A^{(n)}_z$,  taken  from \cite{Abobeih2019},  are tabulated in the appendix.  The nuclear spin expectation values are read out by applying a combination of nuclear-nuclear and electron-nuclear gates, and subsequently reading out the electron spin state as detailed in \cite{TimeCrystal2021}.  As in previous work, \cite{TimeCrystal2021}, we systematically correct for pulse errors and amplitude damping during the readout pulse sequences,  in order to get a best estimate for the spin expectation values.

\subsection{Theoretical methods: Floquet methods} 
We analyse the coherent dynamics with Floquet theory,  a general framework for periodically driven physical systems that has found wide applicability, ranging from  NMR continuous driving \cite{FloquetNMR} to also pulse-based control of NV centers \cite{FloSpec} .  However, Floquet theory 
encompasses several different analytical tools.  Floquet engineering (FE), \cite{Shirley1965} where a system driven by a typically strong or high frequency (non-resonant) field can be shown to correspond to an effective,  static Hamiltonian with renormalised parameters, by averaging over the period of the driving.  Varying the {\em amplitude} of the non-resonant drive, one may tune over the effective Hamiltonian
to polarise the bath \cite{PolChicago}.  A common and widely used approach is the Fourier series decomposition of the one-period Hamiltonian,  in a suitable rotating frame,  which has also been employed for NV pulse-based control of a nuclear-spin bath \cite{FloApp2017}.   Floquet spectroscopy   \cite{FloSpec}  has been introduced in this context:  resonances for pulse-based protocols were shown to correspond to avoided crossings of the underlying Floquet quasi-energies of the pulse-protocol unitary.  Thus the morphology of these single or multiple avoided crossings has proved insightful for analysis of NV-nuclear entanglement and polarisation in terms of Landau-Zener dynamics \cite{FloAdiab2022}.

Here we employed both Fourier analysis and Floquet spectroscopy to analyse our results.\\
 {\em Floquet spectroscopy:} for a system with a temporally periodic Hamiltonian,  ${\hat H}(t+T)={\hat H}(t)$,  Floquet's theorem allows one to write solutions of  the Schr\"odinger equation 
in terms  of quasi-energy states  $|\psi_l (t)\rangle= \exp{(-i{\epsilon}_l t)}   |\Phi_{l}\rangle$
where ${\epsilon}_l$ is the quasi energy.  $|\Phi_l(t) \rangle=|\Phi_l(t+T)\rangle$, $T$ is the period while  $l=1,..,D$ ($D$ is the dimension of the state space).  

One may also obtain eigenstates of the one-period unitary evolution operator ${\hat U}(T) \equiv {\hat U}(T,0)$. The Floquet states $ |\Phi_l\rangle$, obey the eigenvalue equation:
\begin{equation}
{\hat U}(T)|\Phi_l\rangle = \lambda_l  |\Phi_l\rangle \equiv  \exp{(-i{\mathcal E}_l)}   |\Phi_l\rangle
\label{Floquet}
\end{equation}
 where  ${\mathcal E}_l(T)  \equiv \tan^{-1}{ \operatorname{Im}{\lambda_l}/\operatorname{Re}{\lambda_l} }$ is the eigenphase (the Floquet phase).  
 For Floquet spectroscopy numerics,  we diagonalise the full state space of the NV plus a cluster of $N_{nuc} \sim 1-7$ nuclear spins.  Thus we can readily calculate and plot  ${\mathcal E}_l(T)$ as a function of period $T$, to investigate resonances and gain insight on the role played here by overlapping avoided crossings.  
 
 For Fourier series analysis,  a transformation to the toggling frame (the frame of the pulses, see Appendix for details) is widely used,  including for analysis of polarisation protocols \cite{PulsePol,PolCPMG} and their resonances; a key step is to average over a single period.   However,  in order to understand experimental traces as a function of $\tau$,    we must in addition consider also off-resonant behavior (away from $\tau =\tau_r$), as shown below.

\section{Results}
\subsection{Single  spin polarisation: off-resonant behavior}

Details of our analysis are given in the Appendix and here we summarise the key steps.
Away from $\tau=\tau_r$, we introduce a small nuclear detuning,  slightly altering the PulsePol Hamiltonian to:
\begin{equation}
{\hat H} \equiv \sum^{N_{nuc}}_{n=1} g^{(n)} ( {\hat S}_+{\hat I}^{(n)}_- + {\hat S}_-{\hat I}^{(n)}_+)
  + (\omega_I^{(n)} - \omega) {\hat I}^{(n)}_z
\label{ClusterHam}
\end{equation}
where the detuning of each nucleus corresponds to $\delta_n(\tau) = \omega_I^{(n)} - \omega \ll \omega_L$ 
and the protocol frequency $\omega=6\pi/T=3\pi/(2\tau)$.  The resonant nuclear precession frequency is 
$\omega_I^{(n)}=\sqrt{(\omega_L-A^{(n)}_z/2)^2 +(A_x^{(n)}/2)^2}$.
For coherent evolution over $N_p$ pulses,  we can readily show that the polarisation $2\langle {\hat I}^{(n)}_z \rangle$ of a single nuclear spin,  for moderate detuning,  takes the simple form:
\begin{equation}
\mathcal{P}(N_p T) = \left(\frac{2g}{\Omega_r}\right)^2\sin^2\left(\frac{\Omega_r N_p T}{2}\right)
\label{Single_spin_fidelity}
\end{equation} 
where the generalised Rabi frequency $\Omega_r = \sqrt{\delta^2 + (2g)^2}$.  Hence, the maximum population transfer into this state is $\mathcal{P}_{max} = 1/(1 + (\delta/2g)^2)$ at the integer closest to the pulse number $N_p = 2\pi/(\Omega_r T)$.  At resonance,   $\delta = 0$ and  the maximum saturation $\mathcal{P}_{max} = 1$ and the Rabi frequency is $\Omega_r = 2g$.
For $N_p$ greater than this maximal value,  the polarisation oscillates cyclically. Here, by convention $\mathcal{P} \in [-1,1]$. We adopt $\mathcal{P} \in [-1/2,1/2]$, where results can be recovered with the appropriate rescaling.

{\em Asymptotic behavior:} we note the above result is for a single repetition,  $R=1$,  and experimental results range from $R\sim 10^2-10^4$.  One may show that single spin polarisation behaviour tends to an asymptotic envelope in $R \to \infty $ limit.  This is illustrated in Fig.\ref{Fig1} (right panel, red solid line).  Stronger-coupled spins attain the asymptotic form after a few repetitions.  For some weaker coupled blocked spins,  simulations indicate that even $R=10000$ may be insufficient to reach the asymptotic limit.  A notable feature of the polarisation traces is that they exhibit sharp `dips'  at period $T=T_{dip}$,  seen in Fig.\ref{Fig1} (red solid line) and  also seen in the experiments. Here we show that these dips (see Appendix for further details) 
 occur for:
\begin{equation}
T_{\mathrm{dip}} \simeq \frac{T_{r}}{(1 + \mu^2)}\left[1 \pm \frac{n}{3N_p}\sqrt{1 + \mu^2\left(\frac{9N_p^2}{n^2} - 1\right)}\right] \label{side_dip}
\end{equation} 
where $n \in \mathbb{Z}^+$, $n > 0$, $T_r = 6\pi/\omega_I$ and $\mu = 2g/\omega_I \ll 1$
thus $T_{\mathrm{dip}} \simeq T_r\left[1 \pm \frac{n}{3N_p}\right]$.  The experimental traces also contain additional fine-structure due to dephasing arising from the instrumental waiting time $\sim 10\,\mu\mathrm{s}$ in between repetitions.

\begin{figure}[ht!]
\centering
\includegraphics[width=3.4in]{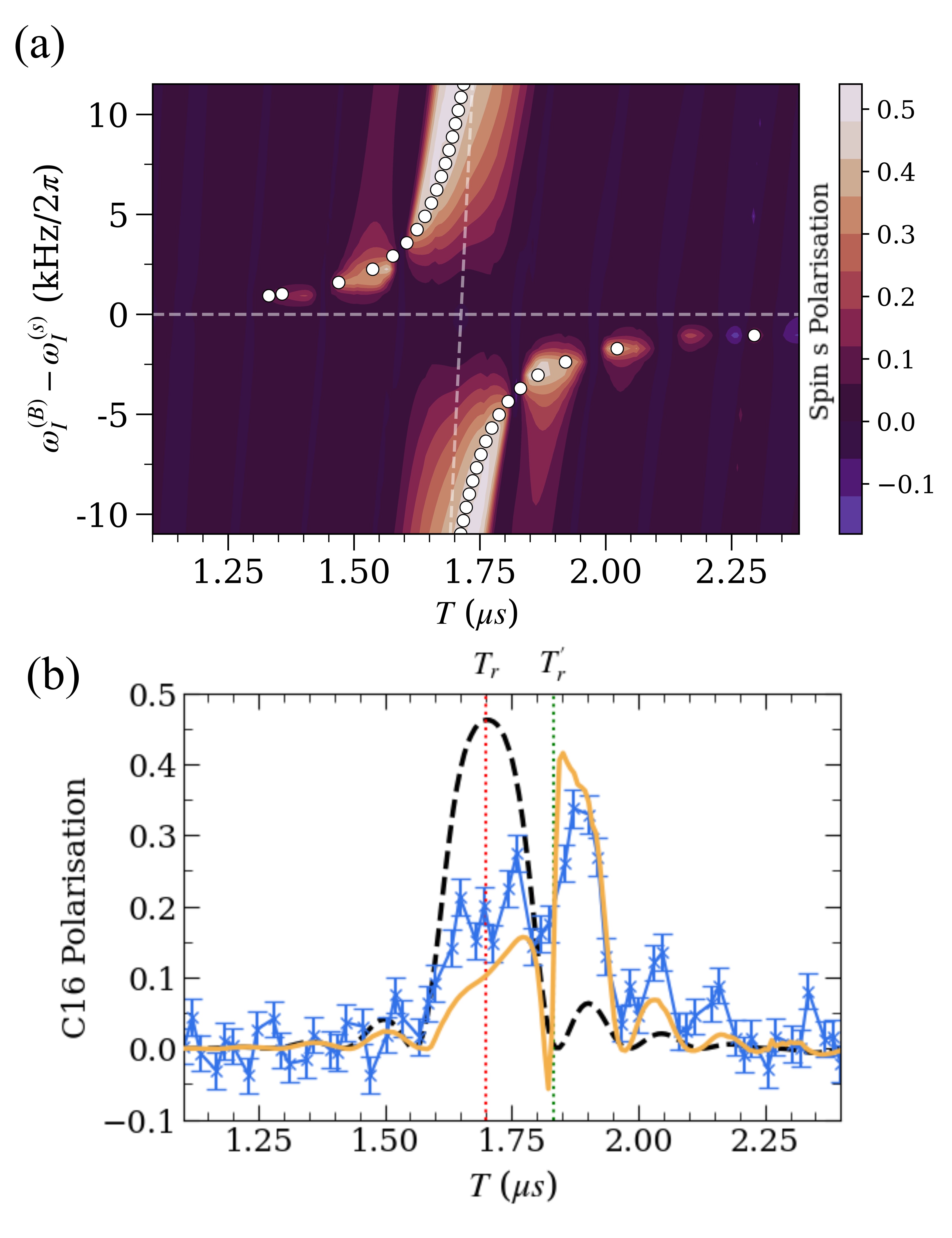}
\caption{  Upper panel tests the analytical expression for the position of the resonance Eq.\ref{Displ} against numerical simulations.  Simulations for this colour map use C3 as a blockade spin and spin with $A_x/(2\pi) = 5$ kHz and a variable $A_z$.  Although agreement with the displaced peak is not exact,  the expression tracks the displacement
quite well.   The lower panel compares experimental polarisation of C16 (blue) in the presence of blockade spin C3,  demonstrating good agreement with simulation (orange) of C3 and C16.  For comparison,  the undisplaced single C16 simulation is shown (black dash line).  Eq.\ref{Displ} is shown to give reasonable agreement with the displaced peak position.  All simulations and experiments in this figure use parameters $N_p = 8$, $R = 100$ which results in a displaced resonance rather than the `wedge' profile obtained for $N_p=4$.}
 \label{Fig3}
\end{figure}

\subsection{Blockade spins: theory and  experiment}

 If an NV center has a proximate $^{13}$C nuclear spin,  at relative orientation such that the spin will have $A^B_{z}\sim 0$ but reasonably strong off-diagonal coupling $A^B_{x}$, we label this spin (see Fig.\ref{Fig1}(a))with superscript $B$ as this `blockade' spin in effect expels the resonances of weakly-coupled spins (with $A^s_{z} \sim 0$ with $s=1,2...$) from the region around $T_r=4\tau_r^{(B)}$,  the expected resonance.  The resonance of the more strongly coupled blocking spin is unperturbed and remains at $T=T_r$.  
The effect was  illustrated in Fig.\ref{Fig1}.   

Although full numerical simulation of clusters of 5-8 nuclei is feasible,  for insight,  our
analysis of spin blockade requires consideration of the NV electron spin as well as the nuclear spin {\em pair}  comprising B and one more nuclear spin.
Pair dynamics involves consideration of an 8-state space.  However,  two states are largely decoupled and analysis reduces to two triplets of coupled states (numerics involve full diagonalisation but,  for insight,  a simpler model is analysed).  

From the Floquet spectroscopy,  this scenario corresponds to  two sets of avoided crossings in the Floquet eigenphases.   This is illustrated in the upper panels of
Fig.\ref{Fig2}.   It shows the pair of avoided crossings: a very broad crossing of width $ \sim A^B_{x}$  for the case of a strong-coupled spin and,  within it,  a very narrow crossing due to the weaker coupled spin since $A^B_{x}\gg A^s_{x}$.  In other words  the  strong avoided crossing involves a pair of states which mostly overlap with states of the single-spin avoided crossing of spin $B$; while the overlapping narrow crossing  involves  states that mostly overlap with the weaker spin states.  

The  lower panels
show the corresponding experimental profiles  for  the weakly coupled nuclear spins C21 with $(A_x,A_z)\equiv (\approx 5.0, -9.7)$ kHz$\times 2\pi$  and C16,  with  $(A_x,A_z)\equiv (5.3, -19.8)$ kHz$\times 2\pi $  respectively.  Their resonances are displaced by the stronger blockade spin spin C3 which has experimentally measured couplings $(A^B_{z}, A^B_{x})=(-11,-59) $ kHz$\times 2\pi$.
There is excellent agreement with simulations. As $N_p=4$ the resonance takes a characteristic wedge shape whereas for $N_p=8$ it is predicted to be fully displaced.

A striking result is that while the C21 resonance is displaced to lower $T$, for the C16 resonance, the converse is true.
Analysis of the 3 state matrix and its eigenvalues gives the position of the weak spin resonance and magnitude of the displacement (see Appendix for derivation):
\begin{equation}
\Delta T_r /T_r \simeq -\frac{{(A_x^{B}})^2}{\omega_I^{(s)}(\omega_I^{(B)} - \omega_I^{(s)})}
\label{Displ}
\end{equation}

Eq.\ref{Displ} is a key result of this work.   Fig.\ref{Fig3} tests this expression against numerics and experimental data.   In the upper panel, the numerical colour map shows  polarisation as a function of $T$ and  detuning $\omega_I^{(B)} - \omega_I^{(s)}$.  The overlaid white dots are from Eq.\ref{Displ} and demonstrate that it provides a robust estimate of the magnitude of the displacement of the polarisation resonance peak.  The lower panel illustrates an  example of the displacement for $N_p=8$ and spin C16.

In contrast to the observed resonance displacement $\Delta T_r$,   the Rabi frequency,  or width of the avoided Floquet crossing,  is  not strongly affected by the blockade spin provided that $|\omega_I^{(B)} - \omega_I^{(s)}|/A_x^{B} \ll 1$ (see appendix for details). 
In general  the measured polarisation is not very significantly reduced,  but  rather is shifted to a 
different period $T_r + \Delta T_r$.   There are particular exceptions,  such as the case of experimental data for a spin simultaneously perturbed by two blockade spins (discussed in the appendix).

\begin{figure}[ht!] 
\includegraphics[width=2.7in]{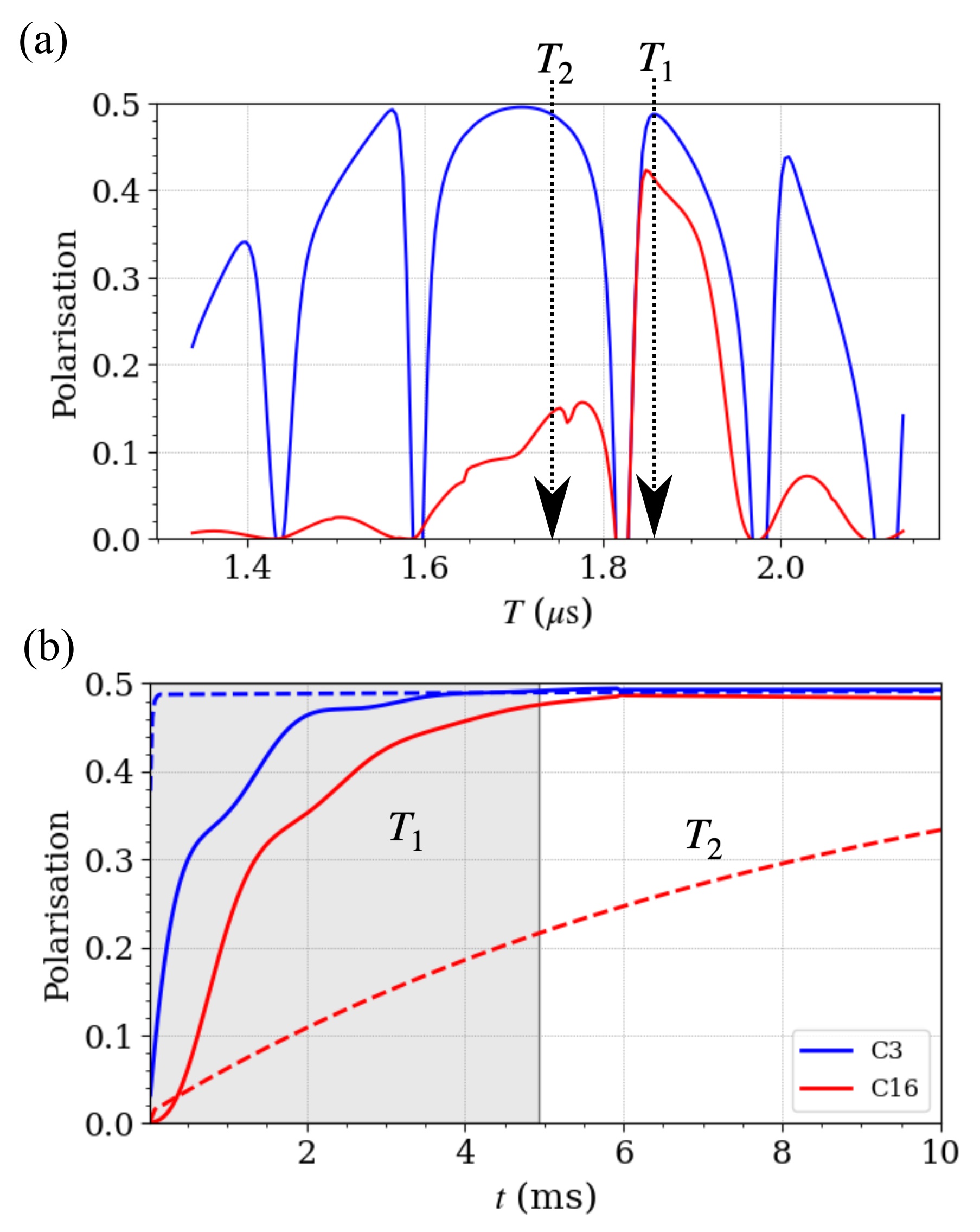}
\caption{Compares the conventional polarisation method of applying PulsePol at a constant $T$ (or $\tau$) to our proposed adaptation of applying two different regimes of $T$ in the presence of blocking spin C3. (a) Simulated polarisation against periodicity of PulsePol, $T$, with parameters $Np = 8$, $R = 100$ of both C3 in blue and C16 with  in red. The two regimes are highlighted as $T_1 \simeq T_r + \Delta T_r\simeq 1.85$ $\mu$s which is a periodicity at the displaced resonance and $T_2 \simeq T_r \simeq 1.74$ $\mu$s near the original resonance. (b) Shows simulations polarisation of both spins with total time. The solid line is the application of 200 repetitions (grey region) at $T_1$ followed by 200 repetitions at $T_2$, the dashed line is the application of 400 repetitions at $T_2$.  A higher level of polarisation in less time is achieved for C16.  For the polarisation of C3, although polarisation rises rapidly for both values of $T$,  driving close to its resonant value at $T=T_2$,  rather than far off resonance at $T=T_1$ is important for ensuring robust polarisation,  to the 0.5 limit.} 
\label{Fig4}
 \end{figure}

\section{Discussion}
 As natural diamond contains of order 1.1\% of $^{13}$C,  we estimate that of order 20\% of NV defects will have a nuclear spin with reasonably strong $A_x$ coupling but with $A_z \sim 0$,  thus is able to produce a blocking effect on distant,  weakly coupled nuclear spins.  
 
 However,  Eq.\ref{Displ} makes clear that the blocking effect is more generic  and will occur wherever a weaker coupled spin is near-degenerate with a stronger coupled spin,  thus  can occur for arbitrary $A_x, A_z$,  provided $A^{(B)}_x \gg A_x$
 and $A^{(B)}_z \sim A_z$.  Thus it should be a relatively common feature in such studies,  and spin-blocking is identifiable via its distinctive spectral profiles,  such as the `wedge' shape for $N_p=4$.  The scenario of two blocking spins acting simultaneously on a weaker spin is less common; but in the present data, we observed the case where two blocking spins 
 act to provide displacements of opposite signs (presented in the Appendix). The result is a sort of destructive cancellation that fully suppresses the resonance peak of the weaker spin.
While the present study considered PulsePol,  our simulations show that similar behavior occurs also for PolCPMG.

 Understanding of the blocking spin mechanism allows one to propose approaches to improve polarisation of weak spins.
Figure 4 demonstrates a method for drastically  improving  the polarisation efficiency  by employing two different $T$.  The upper panel highlights the shifted resonance of spin C16 in the presence of blockade spin C3.  First,  PulsePol is applied with $T \simeq T_r + \Delta T_r$ in the region with the shifted resonance to  polarise weak spins.  Following this, PulsePol with $T \simeq T_r$ is applied to maximise polarisation:  the asymptotic, saturated polarisation  is maximal for $T \simeq T_r$.  The lower panel compares the effectiveness of the two $T$ method compared to the standard  technique of applying the protocol at $T=T_r$ only.  The two spin system of C16 (weak spin) and C3 (blocking spin) was used.  The initial stage (grey region) with $T \simeq T_r + \Delta T_r$,  shows rapid polarisation of both spins (solid lines); the second stage with  $T \simeq T_r$ then yields an improvement on overall polarisation relative to the single $T$,  on-resonance polarisation (dashed line).   

Spin blocking effects are  relevant to  DNP  of  $^{13}$C in the diamond crystal and,  potentially,  external nuclear spins   as well.  Even for different spin species with different Larmor frequencies accidental resonances \cite{Loretz2015} such as those that occur between harmonics for $^1$H and $^{13}$C  might come into play,  but this has not been investigated here.

\subsubsection*{Spin blocking versus dark modes.}
 {\em Dark-bright} modes are a ubiquitous effect occurring in the physics  of 3-level systems \cite{DB1,DB2}:  if we consider two degenerate modes,  with eigenvalues $\epsilon_{1,2} \approx \epsilon$,  independently interacting with a third mode $\epsilon_{3} \sim \epsilon$ with finite coupling strengths  
$g_{1}= g_{2}\equiv g$,  but not with each other,  then  modes $1,2$  hybridize such that one of the hybrid modes fully decouples from mode 3 (dark mode,  effective $g=0$) while the others acquire  enhanced coupling $\sqrt{2}g$ (bright modes).  The spectral signature is generic \cite{DB1}: instead of two independent avoided crossings of width $g_1$ and $g_2$,    the mixing/hybridisation produces a a single wider avoided crossing of width $\sqrt{2}g$ and a completely decoupled state. The role of dark states in impeding polarisation has been noted \cite{TimeCrystal2021} and investigated in a many body context \cite{PolChicago}.

{\em Spin blocking} is a distinct polarisation suppression mechanism.   It occurs for similar regimes,  but for the case where the coupling is highly anisotropic $g_{1} \gg g_{2}$.  The  spectral signature is also generic.  There are once again two separate crossings,  of width not far from the unperturbed widths $g_1$ and $\sim g_2$,   like the unhybridised case.  However only mode 1 remains at $\epsilon \simeq \epsilon_1$.  The weaker coupled mode has its crossing pushed out of the $\epsilon \approx \epsilon_{1,2}$ region.  For the polarisation protocols,  the weaker spin in effect is pushed off-resonance,  which is different from having its effective coupling suppressed, as is the case for a dark state.\\

Off-resonant driving in general does not fully suppress polarisation,  and in principle,  all spins should eventually tend to the $R\to \infty$ limit, which is only maximal on-resonance.  However it may make it extremely inefficient and slow,  potentially allowing imprecisions and decoherent processes to perturb the protocol in a real experiment.  However, unlike dark-mode suppression,  the effect may be mitigated by adjusting the protocol period to account for the shifted resonance.

\section{Conclusions}

In conclusion:  in the present work we introduce,  and  theoretically and experimentally investigate,  the spin-blockade effect,  thus named in analogy to  blockade effects \cite{blockade2009} encountered in other fields of physics.  We show that nuclear spins with strong interactions to the central spin can block polarisation transfer by detuning weaker-coupled spins away from resonance.  This  many-body effect,  detrimental for polarisation efficiency,  can be mitigated by pulse sequences that are tailored to the microscopic configuration of the spin system.  Polarisation transfer to a complex spin system can be highly dependent on the microscopic configuration of the spins and our results thus provide an opportunity for optimization of dynamical nuclear polarisation in various settings.

{\bf Acknowledgements} 
Oliver Whaites acknowledges support from an EPSRC DTP  grant. This work was supported by the Netherlands Organisation for Scientific Research (NWO/OCW) through a Vidi grant. This project has received funding from the European Research Council (ERC) under the European Union's Horizon 2020 research and innovation programme (grant agreement No. 852410).  This work is part of the research programme NWA-ORC (NWA.1160.18.208 and NWA.1292.19.194), (partly) financed by the Dutch Research Council (NWO). This work was supported by the Dutch National Growth Fund (NGF), as part of the Quantum Delta NL programme.
We acknowledge funding from the Dutch Research Council (NWO) through the project ``QuTech Phase II funding: Quantum Technology for Computing and Communication'' (Project No. 601.QT.001).
B.P. acknowledges financial support through a Horizon 2020 Marie Sklodowska-Curie Actions global fellowship (COHESiV, Project Number: 840968) from the European Commission.  We thank M.Markham and D.J.Twitchen from Element Six for providing the diamond.

\section{Appendix}

\subsection{Nuclear spin parameters}

\begin{table}[ht]
\centering
\caption{List of parallel and perpendicular couplings between the $^{13}\mathrm{C}$ spins and the NV. In \cite{Abobeih2019} the  same spins are labelled under a different numbering; the numbering system is included in the table for consistency.}
\begin{tabular}{c | c c c c}
Label & $A_z$/$2\pi$ (kHz) & $A_x$/$2\pi$ (kHz) & $\omega_I^{(i)}$ (rad/$\mu$s) & M Label\\ \hline\hline
C0 & 213.153 & 3 & 2.04 & C9 \\ 
C1 & -36.308 & 26.62 & 2.83 & C18 \\ 
C2 & 20.569 & 41.51 & 2.65 & C12\\ 
C3 & -11.346 & 59.21 & 2.75 & C5 \\ 
C4 & 8.029 & 21.0 & 2.69 & C13\\ 
C5 & 24.399 & 24.81 & 2.64 & C19\\
C6 & -48.58 & 9.0 & 2.86 & C6\\ 
C7 & 14.58 & 10 & 2.67 & C11\\ 
C8 & 7.683 & 4 & 2.69 & C22\\ 
C9 & -20.72 & 12 & 2.78 & C1\\ 
C10 & -23.22 & 13 & 2.78 & C2\\ 
C11 & -13.961 & 9 & 2.75 & C15\\ 
C12 & -31.25 & 8 & 2.81 & C3\\ 
C13 & -14.07 & 13 & 2.76 & C4\\ 
C15 & -5.62 & 5 & 2.73 & C17\\ 
C16 & -19.815 & 5.3 & 2.77 & C14\\
C17 & -4.66 & 7 & 2.73 & C16\\
C18 & 17.643 & 8.6 & 2.66 & C10\\
C20 & -8.32 & 3 & 2.74 & C7\\
C21 & -9.79 & 5.0 & 2.74 & C8\\
C22 & 1.212 & 13 & 2.71 & C21\\
C23 & 2.69 & 11 & 2.70 & C20\\
C24 & -3.177 & 2 & 2.72 & C23\\
C25 & -4.039 & 0.5 & 2.72 & C27\\ 
C26 & -4.225 & 0.771 & 2.72 & C24\\
C27 & -3.873 & 1.247 & 2.72 & C25\\
C28 & -3.618 & 9.472 & 2.72 & C26\\
\end{tabular}
\label{couplings}
\end{table}
\subsection{PulsePol polarisation protocol}
In this section we briefly review the well-studied PulsePol protocol.
Consider a Nitrogen Vacancy (NV) defect in diamond coupled to $N_{nuc}$ $^{13}\mathrm{C}$ spins in a global magnetic field, $B_0$. Under a microwave control field,  the pure dephasing  Hamiltonian in the rotating frame of the NV is:

\begin{equation}
\hat{H}(t) = \left[\omega_L\hat{I}_z^{(n)} + \hat{S}_z(A^{(n)}_z\hat{I}_z^{(n)} + A_x^{(n)}\hat{I}_x^{(n)})\right] + \Omega(t)\hat{S}_{\varphi(t)}
\end{equation}
for a single ($n$-th) nuclear spin and
where $\omega_L = -\gamma_{C13}B_0$ with a gyromagnetic ratio for $^{13}\mathrm{C}$ $\gamma_{C13}$, $A^{(n)}_i$ are $i$th components of the hyperfine coupling strength relative to the $z$-axis between the $n$th nuclei to the NV, $\Omega(t)$ is the waveform of the microwave control field,  $\hat{S}_i$ are in the qubit basis $m_s = \{0,-1\}$, $\hat{I}^{(n)}_i$ are the $n$th nuclear spin-1/2 operators. and $\hat{S}_{\varphi(t)} = \cos(\varphi(t))\hat{S}_x + \sin(\varphi(t))\hat{S}_y$. The experimental external magnetic field strength is aligned with the NV $z$-axis and measured to be $B_0 = 403 \,\mathrm{ G}$. The derivation of this Hamiltonian has assumed the rotating wave approximation. 
For brevity,  we assume $\hbar=1$ in this appendix,  unless considering experimental results in SI units.

The waveform of the microwave control field is dependent on the dynamical decoupling (DD) protocol that is applied. PulsePol is a  DD protocol used for dynamic nuclear polarisation (DNP), as it is state selective.  The form of this pulse protocol is
$\left[\left(\frac{\pi}{2}\right)_{Y} \left({\frac{\tau}{2}}\right) \left(\pi\right)_{-X}  \left({\frac{\tau}{2}}\right)\left(\frac{\pi}{2}\right)_{Y} \left(\frac{\pi}{2}\right)_{X}  \left({\frac{\tau}{2}}\right) \left(\pi\right)_{Y}   \left({\frac{\tau}{2}}\right)\left(\frac{\pi}{2}\right)_{X}\right]^{2N_p} $
where $(\theta)_\varphi$ is a pulse with duration $T_p = \theta/\Omega$ and phase $\varphi$ and $\rightarrow_t$ is free evolution for duration $t$. The Rabi frequency due to the microwave drive is denoted $\Omega$. This protocol is periodic with period $T = 4\tau$ and it is applied for a total time of $T_{tot} = 4N_p\tau$. Using the waveform of the DD control field, the system can be transformed into the frame of the microwave control field, also known as the toggling frame, such that the Hamiltonian in this frame is

\begin{equation}
\hat{H}(t) = \left[\omega^{(n)}_I\hat{I}_z^{(n)} + (f_1(t)\hat{S}_x + f_2(t)\hat{S}_y)(A^{(n)}_z\hat{I}_z^{(n)} + A_x^{(n)}\hat{I}_x^{(n)})\right]
\end{equation}

where $f_i(t)$ are known as the modulation functions with period $f_i(t + T) = f_i(t)$ and are step functions of the form 

\begin{equation}
f_1(t) = \begin{cases}
1 \text{ for } 0 < t < \tau/2 \text{ and } 5\tau/2 < t < 3\tau\\
-1 \text{ for } \tau/2 < t < \tau \text{ and } 2\tau < t < 5\tau/2\\
0 \text{ otherwise }
\end{cases}
\end{equation}

and 
 
\begin{equation}
f_2(t) = \begin{cases}
1 \text{ for } \tau/2 < t < 3\tau/2 \text{ and } 7\tau/2 < t < 4\tau\\
-1 \text{ for } 3\tau/2 < t < 2\tau \text{ and } 3\tau < t < 7\tau/2\\
0 \text{ otherwise }
\end{cases}
\end{equation}

These functions are periodic and so a Fourier decomposition can be used, where for $f_1(t)$

\begin{equation}
f_1(t) = \sum^{\infty}_{k = 0}\left[a_k^{(1)}\cos\left(\frac{k\pi t}{2\tau}\right) + b_k^{(1)}\sin\left(\frac{k\pi t}{2\tau}\right)\right]
\end{equation}

where 

\begin{eqnarray}
a_k^{(1)} &=& \frac{1}{k\pi}\frac{1 - (-1)^k}{2}\left[4\sin\left(\frac{k\pi}{4}\right) - 2\sin\left(\frac{k\pi}{2}\right)\right]\nonumber \\
b_k^{(1)} &=& \frac{1}{k\pi}\frac{1 - (-1)^k}{2}\left[-4\cos\left(\frac{k\pi}{4}\right) + 2\right]
\end{eqnarray} 

and the remaining modulation function $f_2(t)$ can be found in a similar fashion, only with coefficients $a_k^{(2)} = -(-1)^k b_k ^{(1)}$ and $b_k^{(2)} = (-1)^k a_k^{(1)}$ due to the shift of $t - \tau/2$ in their time dependence. If it is assumed that $A_x << \pi/\tau$, then the time dependence of the Hamiltonian can be resolved in a perturbative fashion using the Magnus expansion. To first order, the evolution operator is $\hat{U}(T) = \exp[-i\hat{H}_{avg} T]$, where $\hat{H}_{avg}$ is the time period averaged Hamiltonian constructed as 

\begin{equation}
\hat{H}_{avg} = \frac{1}{T}\int_0^T \hat{H}(t) \,dt
\end{equation} 

hence removing the time dependence in place of time averaged field strengths. This assumption of small times scales is fulfilled by transforming into the interaction frame of the nuclear spin of interest and assuming that $A_x \ll 1/T$. Choosing the resonance $\tau_r = 3\pi/(2\omega_L)$ or $k = 3$, and taking the time average for the first order perturbation, the first order Hamiltonian can be found to be:

\begin{equation}
\hat{H}_{avg} = \sum_{n = 0}^{N_{nuc}}\left[g^{(n)}(\hat{S}_+\hat{I}_-^{(n)} + \hat{S}_-\hat{I}_+^{(n)}) + \delta^{(n)}\hat{I}_z^{(n)}\right]\label{avg_ham}
\end{equation}
where we assume the Hamiltonian is an independent sum of $N_{nuc}$ spins, with all interactions mediated by the central electronic spin.
We define $g^{(n)} = A_x^{(n)}\alpha/4$, $\alpha = 2(\sqrt{2} + 2)/3\pi$ and $\delta^{(n)} = \omega^{(n)}_I - \omega_p$ ($\simeq A_z^{(n)}/2$ for $\omega_p = 6\pi/T \simeq \omega_L$) and resonance harmonic $k$.  For simplicity, as stated previously $k = 3$  is chosen and will be absorbed into the relationship between protocol frequency and period, such that $\omega_p = 2k\pi/T = 6\pi/T$. The nuclear precession frequency is $\omega^{(n)}_I = \sqrt{(\omega_L - A_z^{(n)}/2)^2 + (A_x^{(n)}/2)^2}$. This Hamiltonian is appropriate for small detuning from the nuclear precession frequency, or that $\delta \ll \omega_L$. If this is not satisfied, higher order terms may be needed.

\section{Single Spin Polarisation}

We now review the simplest NV plus single nuclear spin case.  While this is well-known, it is important for our study to keep track of the detuning $\delta$ as we are interested in the behaviour away from the resonant pulse period.  Equation \ref{avg_ham},  for the case of a single spin, takes the matrix form:

\begin{equation}
\hat{H}_{avg} =  \begin{matrix}
|\uparrow\uparrow\rangle\\
|\downarrow\uparrow\rangle\\
|\uparrow\downarrow\rangle\\
|\downarrow\downarrow\rangle
\end{matrix}
\begin{pmatrix}
\delta/2 & 0 & 0 & 0 \\
0 & \delta/2 & g & 0 \\
0 & g & -\delta/2 & 0 \\
0 & 0 & 0 & -\delta/2
\end{pmatrix}
\label{single_H_mat}
\end{equation} where the superscript in the previous section is removed in the single spin case for simplicity. For ease of notation, the NV states are re-labelled as $|0/-1\rangle = |\uparrow/\downarrow\rangle$.

From whence we see that the stretched states are decoupled thus:

\begin{eqnarray}
|\chi_1\rangle &=& |\uparrow\uparrow\rangle, \quad \epsilon_1 = \delta/2\\
|\chi_2\rangle &=& |\downarrow\downarrow\rangle, \quad \epsilon_2 = -\delta/2
\end{eqnarray}

while the  anti-aligned states form a 2x2 subspace of $\{|\uparrow\downarrow\rangle,|\downarrow\uparrow\rangle\}$, or a pseudo-spin 1/2 Hamiltonian of the form \begin{equation}
\hat{H} = \delta\hat{I}_z + 2g\hat{I}_x = \mathbf{h}\cdot\hat{\mathbf{I}}
\end{equation}
where $\mathbf{h} = \omega(\sin\theta_p,0,\cos\theta_p)$ with eigenstates and eigenvalues:

\begin{eqnarray}
|\chi_3\rangle &=& \cos\left(\frac{\theta_p}{2}\right)|\downarrow\uparrow\rangle + \sin\left(\frac{\theta_p}{2}\right)|\uparrow\downarrow\rangle, \quad \epsilon_3 = \frac{\omega}{2} \nonumber\\
|\chi_4\rangle &=& \sin\left(\frac{\theta_p}{2}\right)|\downarrow\uparrow\rangle - \cos\left(\frac{\theta_p}{2}\right)|\uparrow\downarrow\rangle, \quad \epsilon_4 = -\frac{\omega}{2}\nonumber\\
\end{eqnarray} 

defining $\omega = \sqrt{\delta^2 + 4g^2}$ and $\tan\theta_p = 2g/\delta$. The detuning from the nuclear resonance creates an effective magnetic field in this subspace which is not aligned with the $x$-axis. This misalignment, means that maximum transfer between these two states is not possible.  To see this, consider an initial state of $|\psi(0)\rangle = |\uparrow\downarrow\rangle$.  With the unitary $\hat{U}(t) = \exp(-i\hat{H}_{av}t)$ and $|\psi(t)\rangle = \hat{U}(t)|\psi(0)\rangle$ we readily obtain:
\begin{eqnarray}
|\psi(T_{tot} = N_pT)\rangle &=& \left[\cos(\varphi(N_pT)) + i\cos(\theta_p)\sin(\varphi(N_pT))\right]\nonumber \\
|\uparrow\downarrow\rangle &-& i\sin(\theta_p)\sin(\varphi(N_pT))|\downarrow\uparrow\rangle
\label{single_spin_trans}
\end{eqnarray}
where $\varphi(N_pT) = \omega N_pT/2$. The population from the $|\uparrow\downarrow\rangle$ state will transfer into the corresponding flip flop state $|\downarrow\uparrow\rangle$. The population in this state is then 

\begin{equation}
P_{\downarrow\uparrow}(N_pT) = |\langle \downarrow\uparrow|\psi(N_pT)\rangle|^2 = \sin^2(\theta_p) \sin^2\left(\frac{\omega N_p T}{2}\right)
\label{Single_spin_fidelity}
\end{equation} 
and hence, the maximum population transfer into this state is $P_{\downarrow\uparrow}^{max} = \sin^2\theta_p = 1/(1 + (\delta/2g)^2)$ at $N_p = 2\pi/(\omega T)$. On resonance, or $\delta = 0$, the maximum saturation $P_{\downarrow\uparrow}^{max} = 1$ as expected.
The system undergoes oscillations between these two states characterised by the frequency $\Omega_r = \omega= \sqrt{\delta^2 + 4g^2}$, the generalised Rabi-frequency.

\subsubsection{Mixed states and repetitions}
 In general,  nuclear spins such as $^{13}\mathrm{C}$ are not naturally in  a pure state but instead are in thermal mixture of equal parts $|\uparrow\rangle$ and $|\downarrow\rangle$. This is often represented as a density matrix, where the thermal mixture is initially 

\begin{equation}
\rho_{I}(0) = \frac{1}{2}
\begin{matrix}
|\uparrow\rangle\\
|\downarrow\rangle
\end{matrix}
\begin{pmatrix}
1 & 0\\
0 & 1
\end{pmatrix} = \frac{\rho_{\uparrow}}{2} + \frac{\rho_\downarrow}{2}.
\end{equation} 

However, the NV is initialised into one of the basis states, where here it is taken that the NV is initialised into the $|\uparrow\rangle$ state such that $\rho_{NV}(0)  = \rho_\uparrow$. As with the state representation, the density matrix of the product space of NV-C13 is constructed through the tensor product $\rho(0) = \rho_{NV}(0)\otimes\rho_I(0)$. The initial density matrix then evolves in time according to the transformation 

\begin{equation}
\rho(N_pT) = \hat{U}(N_pT)\,\rho(0)\,\hat{U}^{\dagger}(N_pT).
\end{equation} 

For this system, the evolved density matrix is found to be

\begin{eqnarray}
&\rho(N_p T) = \frac{1}{2}&\nonumber\\
&\begin{matrix}
|\uparrow\uparrow\rangle\\
|\downarrow\uparrow\rangle\\
|\uparrow\downarrow\rangle\\
|\downarrow\downarrow\rangle
\end{matrix}
\begin{pmatrix}
1 & 0 & 0 & 0\\
0 & |\beta(N_pT)|^2 & \alpha^*(N_pT)\beta(N_pT) & 0\\
0 & \alpha(N_p T)\beta^*(N_p T) & |\alpha(N_p T)|^2 & 0\\
0 & 0 & 0 & 0
\end{pmatrix}&
\label{full_single_density}
\end{eqnarray}

where $\alpha(N_pT) = \cos(\varphi(N_pT)) + i\cos(\theta_p)\sin(\varphi(N_pT))$ and $\beta(N_p T) = - i\sin(\theta_p)\sin(\varphi(N_pT))$. In order to simplify this and explicitly show the evolution of the $^{13}\mathrm{C}$ spin, a partial trace can be performed on this density matrix in order to reduce the 4x4 matrix to a 2x2 matrix and trace out the NV states. A partial trace over one subspace, $\mathcal{H}_B$, with basis states $\{|b_l\rangle\}$ in a dual system between this subspace and $\mathcal{H}_A$ with basis states $\{|a_i\rangle\}$ is defined as 

\begin{equation}
\rho_A = \text{tr}_B(\rho_{AB}) = \sum_{ijlk} c_{ijlk}|a_i\rangle\langle a_j|\langle b_l|b_k\rangle
\label{partial_trace}
\end{equation} 

where the dual space density matrix is decomposed as $\rho_{AB} = \sum_{ijlk} c_{ijlk}|a_i\rangle\langle a_j|\otimes|b_l\rangle\langle b_k|$. By performing this operation over the NV basis states on the full density matrix in Eq.\eqref{full_single_density}, the dynamics of the nuclear spin can be reduced to 

\begin{equation}
\rho_I(N_p T) = \frac{1}{2}\begin{matrix}
|\uparrow\rangle\\
|\downarrow\rangle
\end{matrix}
\begin{pmatrix}
1 + \mathcal{P}(N_p T) & 0\\
0 & 1 - \mathcal{P}(N_p T)
\end{pmatrix}
\label{rho_pol}
\end{equation} 

where $\mathcal{P}(N_p T) = |\beta(N_p T)|^2 = \sin^2(\theta_p)\sin^2\left(\Omega_r N_p T/2\right)$. This density matrix can be represented in the Pauli basis as $\rho(N_pT) = (\mathbb{I} + \mathcal{P}(N_p T)\hat{\sigma}_z)/2$. The polarisation, defined as $\langle\hat{\sigma}_z(t)\rangle/2 = \text{tr}(\rho(t)\hat{\sigma}_z)/2$ is found to be 

	 \begin{figure}[ht!]
\centering
\includegraphics[width=3.7in]{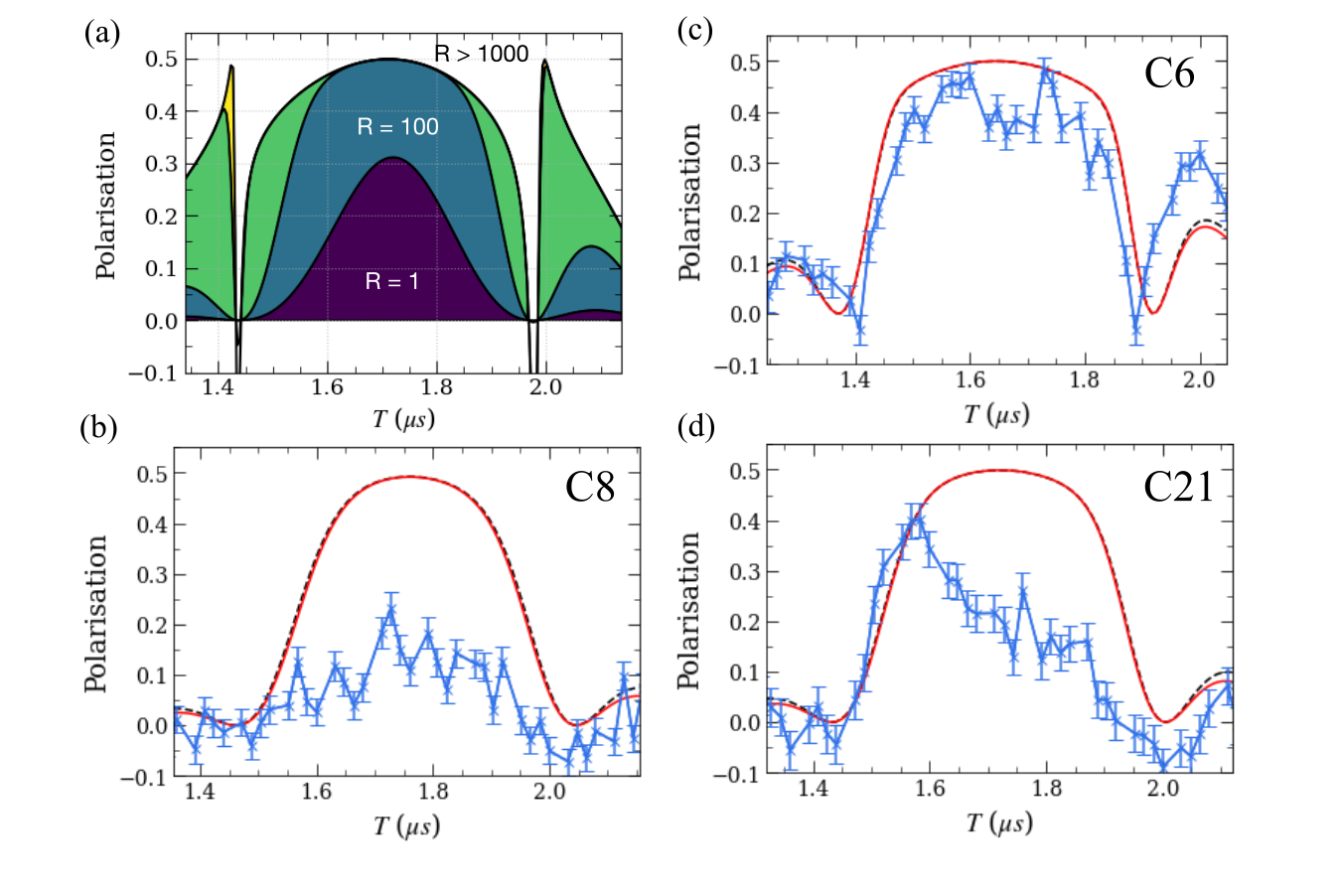}
	\caption{(a) Simulation showing that with increasing repetitions $R$,  of the polarisation sequence, in principle all spins tend to an asymptotic polarisation envelope.  Isolated strong spins attain the $R \to \infty$ limit for $R \sim 10$,  while isolated very weak coupled spins may require $R\gg 1000$.
	Comparisons of single spin simulations of polarisation in (red) to experimental data (blue) for three weak coupled spins in a cluster C6 (b), C8 (c) and C21 (d).  
	The parameters used here are $R = 1000$ and $N_p = 4$.  The modest-coupling strength  C6 is close to asymptotic. The weak coupled spins C8 and C21 are far from the limit, not only because of the weak coupling,  but also because of the many-body spin blocking investigated here. }	
	\label{Asymptote}
\end{figure}

\begin{equation}
\langle\hat{I}_z(N_p T)\rangle = \frac{\mathcal{P}(N_p T)}{2} = \frac{1}{2}\sin^2(\theta_p)\sin^2\left(\frac{\Omega_r N_p T}{2}\right)
\label{pol_single}
\end{equation}
and completely equivalent to the result of the previous section.

However, in the experiment the NV is repeatedly optically re-initialised into the $m=0$ state,  with a repetition number $R$.  All single-spins simulations
saturate to an asymptotic envelope as $R \to \infty$ although the rate at which the limit is approached differs significantly,  as illustrated in Fig.\ref{Asymptote}.
Weaker spins evidently polarise more slowly, even in the isolated spin case.  In addition there are the many-body effects, such as dark state formation and the new `spin-blocking' that we investigated here that can significantly reduce the polarisation rate.

Even in the single-spin case,  both the simulated and experimental polarisation traces show a detailed structure of `polarisation dips'.  In the following sections we analyse a key mechanism which causes sharp polarisation dips, prior to discussing many-spin effects.

\subsubsection{Polarisation envelope side-dips}

In Fig.\ref{Asymptote} and other figures one sees  extremely large `side-dips',  symmetrically distributed about the resonant $T_r$ where the average Hamiltonian model
predicts zero polarisation,  but full numerics allow even negative polarisation.
 
 Within the average Hamiltonian model,  the polarisation in Eq.\eqref{pol_single} is zero when the condition
\begin{equation}
\sin\left(\frac{\Omega_r N_p T}{2}\right) = 0
\end{equation} 

is met. Solving this equation for $T$ yields the position for a series of such side `dips', $T_{\mathrm{dip}}$, to be

\begin{eqnarray}
T_{\mathrm{dip}} & = & \frac{T_{r}}{(1 + \mu^2)}\left[1 \pm \frac{n}{kN_p}\sqrt{1 + \mu^2\left(\frac{k^2N_p^2}{n^2} - 1\right)}\right]\nonumber \\
& \simeq & T_r\left[1 \pm \frac{n}{kN_p}\right]
\label{side_dip}
\end{eqnarray} 
where $n \in \mathbb{Z}^+$, $n > 0$ is the $n$-th side dip.

 $k$ is the resonance harmonic and , $T_r = 2k\pi/\omega_I$ while $\mu = 2g/\omega_I$.  For harmonic $k = 3$,  the side dips are approximately 

\begin{equation}
T_{\mathrm{dip}} \simeq T_r\left(1 \pm \frac{n}{3N_p}\right)
\end{equation}

assuming $\mu \ll 1$ since the nuclear spins and magnetic field used here $g \ll \omega_I$.

\section{Two nuclear spins}
The average Hamiltonian for the two-spin system is a straightforward extension of the single-spin case:

\begin{eqnarray}
\hat{H}_{avg} &=& g_1(\hat{S}_+\hat{I}^{(1)}_- +  \hat{S}_-\hat{I}^{(1)}_+) \nonumber\\
 &+& \delta_1\hat{I}^{(1)}_z + g_2(\hat{S}_+\hat{I}^{(2)}_- +  \hat{S}_-\hat{I}^{(2)}_+) + \delta_2\hat{I}^{(2)}_z
\end{eqnarray} 

and corresponds to a  $8\times8$ matrix that may  be decomposed into 4 independent subspaces.
As with the single spin model, the stretched states $|\uparrow\uparrow\uparrow\rangle$ and $|\downarrow\downarrow\downarrow\rangle$ are eigenstates of $\hat{H}_{avg}$ with 

\begin{eqnarray}
|\chi_1\rangle &=& |\uparrow\uparrow\uparrow\rangle, \quad \epsilon_1 = \frac{\delta_1 + \delta_2}{2} = \frac{\delta_+}{2}\\
|\chi_2\rangle &=& |\downarrow\downarrow\downarrow\rangle, \quad \epsilon_2 = -\frac{\delta_1 + \delta_2}{2} = -\frac{\delta_+}{2}
\end{eqnarray} 

The remaining states form two decoupled subspaces of $\{|\uparrow\downarrow\uparrow\rangle,|\downarrow\uparrow\uparrow\rangle,|\uparrow\uparrow\downarrow\rangle\}$ with overall magnetisation $M_j = 1/2$ and $\{|\downarrow\downarrow\uparrow\rangle,|\uparrow\downarrow\downarrow\rangle,|\downarrow\uparrow\downarrow\rangle\}$ with overall magnetisation $M_j = -1/2$. 
For the case of the  $M_j = 1/2$ subspace,  the  Hamiltonian matrix takes the form:

\begin{equation}
\hat{H}_{avg} = \begin{matrix}
|\uparrow\downarrow\uparrow\rangle\\
|\downarrow\uparrow\uparrow\rangle\\
|\uparrow\uparrow\downarrow\rangle
\end{matrix}
\begin{pmatrix}
-\delta_-/2 & g_1 & 0\\
g_1 & \delta_+/2 & g_2\\
0 & g_2 & \delta_-/2
\end{pmatrix}\label{2_spin_ham}
\end{equation} 

where $\delta_- = \delta_1 - \delta_2$.   

\subsection{Dark and bright states}
For the completely degenerate case 
\begin{equation}
\hat{H}_{avg}  \equiv \delta \hat{I} + \begin{matrix}
|\uparrow\downarrow\uparrow\rangle\\
|\downarrow\uparrow\uparrow\rangle\\
|\uparrow\uparrow\downarrow\rangle
\end{matrix}
\begin{pmatrix}
0 & g_1 & 0\\
g_1 & 0 & g_2\\
0 & g_2 & 0
\end{pmatrix}\label{dark}
\end{equation} 
where $\delta$ is arbitrary and $\hat{I}$  is the $3X3$ identity matrix,   there are well-studied eigenstates:  
\begin{eqnarray}
|\psi_{B_-}\rangle &=& |\uparrow\rangle_{NV}[\cos\varphi|\downarrow\uparrow\rangle + \sin\varphi|\uparrow\downarrow\rangle - |\downarrow\uparrow\uparrow\rangle]\nonumber\\
|\psi_{B_+}\rangle &=& |\uparrow\rangle_{NV}[\cos\varphi|\downarrow\uparrow\rangle + \sin\varphi|\uparrow\downarrow\rangle + |\downarrow\uparrow\uparrow\rangle]\nonumber \\
|\psi_D\rangle &=& |\uparrow\rangle_{NV}[\cos\varphi|\uparrow\downarrow\rangle - \sin\varphi|\downarrow\uparrow\rangle]\label{dark_0}
\end{eqnarray}

where $ |\psi_D\rangle$ is termed a `dark state', with eigenvalue $\epsilon$, and $|\psi_{B_\pm} \rangle $ so-called `bright' states.   $g_{rms} = \sqrt{g_1^2 + g_2^2}$,  with $ \cos\varphi=g_1/g_{rms}$,  $ \sin\varphi=g_2/g_{rms}$

The dark state $ |\psi_D\rangle $ has no overlap with the fully polarised state $ |\downarrow\uparrow\uparrow\rangle $,  thus any state that acquires a significant dark state component may not polarise.  For the bright states,  the converse is true.  
For the case $g_1=g_2=g$ the problem is even simpler as the matrix reduces to $g\hat{S}_x$, the spin 1 angular momentum $x$ matrix.\\

However, we need to consider the behaviour in the presence of detuning, hence away from resonance, as well as spins that are not perfectly degenerate. Neither the uncoupled Zeeman states nor the dark-bright states are eigenstates of the general $\hat{H}_{avg}$ case. 
For insight, we represent $\hat{H}_{avg}$ in  a slightly modified  alternative  `dark-bright' basis $\{|\psi_{pol}\rangle,|\psi_B^{(1)}\rangle,|\psi_D^{(1)}\rangle\}$ where 

\begin{eqnarray}
|\psi_{pol}\rangle &=& |\downarrow\uparrow\uparrow\rangle\\
|\psi_B^{(1)}\rangle &=& |\uparrow\rangle_{NV}[\cos\varphi|\downarrow\uparrow\rangle + \sin\varphi|\uparrow\downarrow\rangle]\label{bright_1}\\
|\psi_D^{(1)}\rangle &=& |\uparrow\rangle_{NV}[\cos\varphi|\uparrow\downarrow\rangle - \sin\varphi|\downarrow\uparrow\rangle]\label{dark_1}
\end{eqnarray} 
for $\tan\varphi = g_2/g_1$.  In this basis,

\begin{equation}
\hat{H}_{av}' = \begin{matrix}
|\psi_{pol}\rangle\\
|\psi_B^{(1)}\rangle\\
|\psi_D^{(1)}\rangle
\end{matrix}
\begin{pmatrix}
\delta_+/2 & g_{rms} & 0\\
g_{rms} & -\delta_-\cos(2\varphi)/2 & \delta_-\sin(2\varphi)/2\\
0 & \delta_-\sin(2\varphi)/2 & \delta_-\cos(2\varphi)/2
\end{pmatrix}
\end{equation} 
   that we use to analyse two important cases.

\subsection{Dark-bright states in the presence of detuning}

First we consider the case where the two nuclear spins are  degenerate  ($A^{(1)}_z= A^{(2)}_z)$ so $\delta_- = 0$ and $\delta_+ = 2\delta_1 = 2\delta$, a detuning which varies as the experiment sweeps $T$ through the resonance $T = T_r$. In practice,  we require $\delta_-/2 \ll g_{rms}$. The dark state $|\psi_D^{(1)}\rangle$ decouples from the other bright states. The other two states form a separate subspace $\{|\psi_{pol}\rangle|\psi_B^{(1)}\rangle\}$, such that the Hamiltonian is $\hat{H} = \delta/2 \mathbb{I} + \delta\hat{I}_z + 2g_{rms}\hat{I}_x$. This Hamiltonian is  similar to the single spin model with $g \to g_{rms}$ and an extra global energy shift.  Hence:

\begin{eqnarray}
|\chi_+^{(1)}\rangle &=&  \cos\left(\frac{\theta}{2}\right)|\psi_{pol}\rangle + \sin\left(\frac{\theta}{2}\right)|\psi_B^{(1)}\rangle , \quad \epsilon^{(1)}_+ = \frac{\delta + \omega}{2}\nonumber\\
|\chi_-^{(1)}\rangle &=&  \sin\left(\frac{\theta}{2}\right)|\psi_{pol}\rangle - \cos\left(\frac{\theta}{2}\right)|\psi_B^{(1)}\rangle , \quad \epsilon^{(1)}_- = \frac{\delta - \omega}{2}\nonumber \\
|\chi_D^{(1)}\rangle &=& |\psi_D^{(1)}\rangle, \quad \epsilon_0 = 0\label{dark_state}
\end{eqnarray} 
where $\tan\theta = 2g_{rms}/\delta$ and $\omega = \sqrt{\delta^2 + 4g_{rms}^2}$ are dependent on the detuning.  The dark state is $\delta-$ independent and is fully decoupled (effective zero coupling to the polarisation state) even for $g_1=g_2$ . As $\delta$ is varied, $|\chi_\pm^{(1)}\rangle$ sweep through a single avoided crossing with enhanced coupling $g_{rms}$.

If we take the initial state of the system to be $|\uparrow\rangle$ for the NV and the nuclear spins in a thermal mixture, thus  $|\psi(0)\rangle = |\uparrow\downarrow\uparrow\rangle$,  temporal evolution will yield 

\begin{eqnarray}
|\psi(N_pT)\rangle &=& e^{-i\delta N_p T/2}\cos\varphi\left[\alpha(N_p T)|\psi_B^{(1)}\rangle + \beta(N_p T)|\psi_{pol}\rangle\right] \nonumber \\
& - & \sin\varphi|\psi_{D}^{(1)}\rangle
\label{2_spin_prop}
\end{eqnarray} 

where $\alpha(t) = \cos(\omega t/2) + i\cos(\theta)\sin(\omega t/2)$, $\beta(t) = -i\sin(\theta)\sin(\omega t/2)$ and $\cos\varphi = \frac{g_1}{g_{rms}}$. 
 This is similar to the previous single spin only with an extra, time independent, term involving the dark state.  The population of the state in the polarised state is then

\begin{equation}
P_{pol}(N_p T) =  \cos^2 \varphi \sin^2\theta  \sin^2\left(\frac{\omega N_p T}{2}\right).
\label{2_spin_transfer}
\end{equation} 

\begin{figure}[t]\centering
		\includegraphics[width = 3.5in]{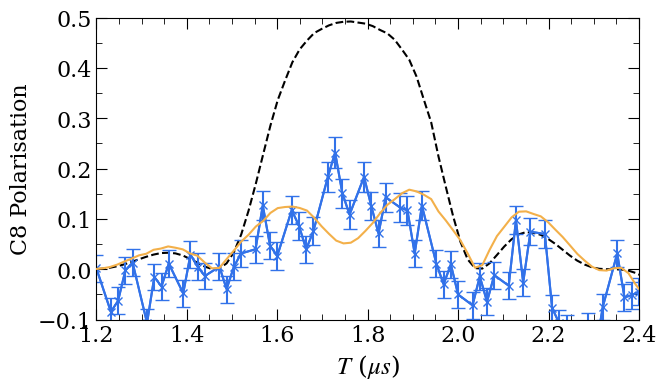}
	\caption{Illustration of polarisation saturation for spin C8 due to spin-pair dark states formed with a nearly degenerate spin C4. Simulations of single spin C8 (dashed line) is shown to polarise fully after $R = 1000$ of PulsePol with $N_p = 4$. This is not seen in experiment (blue) which only reaches a polarisation of ~0.2. This experimental saturation is reproduced to good effect by simulation of the spin-pair C8 and C4 (orange).} 
	\label{C4andC8}
\end{figure} 

This population has the same off-resonance envelope with $\sin\theta$ blocking population transfer if $\delta \neq 0$, but also has an extra term involving $\varphi$, which quantifies overlap with the dark state and suppresses population transfer to the polarised state, as $\cos^2\varphi \leq 1$. This is seen in Eq.\eqref{2_spin_transfer} at arbitrary $\delta(\tau)$. 
An equivalent approach can be used to analyse the  $M_j = -1/2$ subspace $\{|\downarrow\downarrow\uparrow\rangle,|\uparrow\downarrow\downarrow\rangle,|\downarrow\uparrow\downarrow\rangle\}$.\\

In summary,  for   degenerate spins, dark states cause a proportion of the spin's population to remain unpolarised , even for a large number of repetitions $R$.
Perfect degeneracy is not typical in a realistic setting, and in most cases slow mixing between dark and bright states allows slow polarisation.  
An example of dark state polarisation suppression is shown in Fig.\ref{C4andC8} between C8 and C4, which are almost degenerate, as in table.\ref{couplings}. There is a significant reduction in polarisation between the single spin simulation of C8 and a two spin simulation with only C8 and C4,  which better captures the experimental behavior.

\subsection{Spin Blockade}
\begin{figure}[t]\centering
		\includegraphics[width = 3.3in]{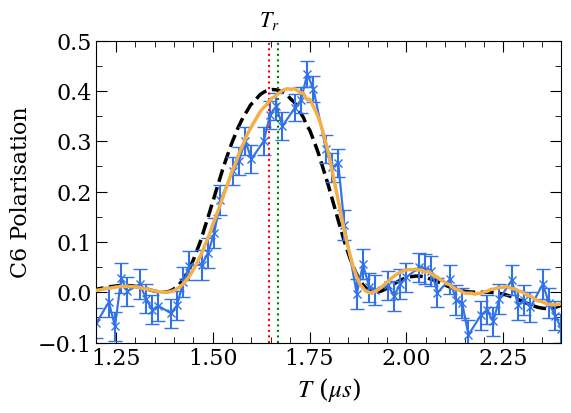}
	\caption{Polarisation of C6 displaced by blocking spin C3. Simulations of C6 polarisation with C3 (orange line) after $R = 100$ of PulsePol with $N_p = 4$ has a slightly displaced polarisation peak from the single C6 spin (dashed black-line) despite having a stronger perpendicular coupling of $A_x = 7.68 \,\mathrm{kHz}/2\pi$. This displacement is seen in experimental data for C6 polarisation (blue).} 
	\label{C6}
\end{figure}

In this section we present the details of the spin blocking effect which is introduced in this work.  We consider the case of anisotropic coupling strengths thus $g_1\gg g_2$.   We relabel $g_1\equiv G$ and $g_2\equiv g$, hence $G\gg g$,  to clearly distinguish the strong from the weak coupling spin.  However,  the spins are not fully degenerate, or $\delta_- \neq 0$,  but $\delta_- \ll G$.
 Such a scenario may be common when using NV centres to polarise a cluster of distant nuclear spins.  There are typically a small number of proximate or near-proximate spins such as spin C3 where this condition is satisfied by a larger number of weaker coupled spins such as C16,C21,C25 etc.  The anisotropy requirement  is not too stringent:  blockade spins can displace the resonance of `local' spins with moderate  perpendicular coupling,  an effect which may need to be considered when using local clusters as memory registers.  Fig.\ref{C6} demonstrates an example: spin  C6 has reasonable coupling $A_x/(2\pi)=9$ kHz, but still experiences a small resonance shift from blockade spin C3,  suggesting that the blocking effect is quite common.

For this case, since we write the effective Hamiltonian in Eq.\eqref{2_spin_ham}  $\hat{H}_{avg} = \hat{H}_0 + \hat{V}$  as a zeroth order Hamiltonian plus perturbation matrix where:

\begin{equation}
\hat{H}_0 + \hat{V} = \begin{matrix}
|\uparrow\downarrow\uparrow\rangle\\
|\downarrow\uparrow\uparrow\rangle\\
|\uparrow\uparrow\downarrow\rangle
\end{matrix}
\begin{pmatrix}
-\delta_-/2 & G & 0\\
G & \delta_+/2 & 0\\
0 & 0 & \delta_-/2
\end{pmatrix} + 
\begin{pmatrix}
0 & 0 & 0\\
0 & 0 & g\\
0 & g & 0
\end{pmatrix}
\end{equation} 
The detunings are relabelled to $\delta_1 \to \delta_B$ and $\delta_2 \to \delta_s$ for a blockade spin and `small' spin respectively. The unperturbed Hamiltonian $\hat{H}_0$ can be reduced into two subspaces. One subspace only contains the state $\{|\uparrow\uparrow\downarrow\rangle\}$ which decouples from the others. Therefore the eigenstate and eigenvalue can be read from the matrix to be $|\chi^{(0)}_3\rangle = |\uparrow\uparrow\downarrow\rangle, \quad \epsilon_3^{(0)} = \delta_-/2$. The remaining states form their own subspace  $\{|\uparrow\downarrow\uparrow\rangle,|\downarrow\uparrow\uparrow\rangle\}$ with Hamiltonian 

\begin{equation}
\hat{\tilde{H}}_0 = \begin{matrix}
|\downarrow\uparrow\uparrow\rangle\\
|\uparrow\downarrow\uparrow\rangle
\end{matrix}
\begin{pmatrix}
\delta_+/2 & G\\
G & -\delta_-/2
\end{pmatrix} = \frac{\delta_s}{2}\mathbb{I} + \delta_B\hat{I}_z + 2G\hat{I}_x\\
\end{equation}

As expected,  this takes the same form as the single spin case with an extra global energy term from the second spin. The corresponding eigenstates and eigenvalues are:

\begin{eqnarray}
|\chi^{(0)}_1\rangle &=& \cos\left(\frac{\theta_p}{2}\right)|\downarrow\uparrow\uparrow\rangle + \sin\left(\frac{\theta_p}{2}\right)|\uparrow\downarrow\uparrow\rangle, \quad \epsilon^{(0)}_1 = \frac{\delta_s + \omega}{2}\nonumber\\
|\chi^{(0)}_2\rangle &=& \sin\left(\frac{\theta_p}{2}\right)|\downarrow\uparrow\uparrow\rangle - \cos\left(\frac{\theta_p}{2}\right)|\uparrow\downarrow\uparrow\rangle, \quad \epsilon^{(0)}_2 = \frac{\delta_s - \omega}{2}\nonumber\\
|\chi^{(0)}_3\rangle &=& |\uparrow\uparrow\downarrow\rangle, \quad \epsilon^{(0)}_3 = \frac{\delta_-}{2}\label{chi_B_3}
\end{eqnarray} 

where  $\omega = \sqrt{\delta_B^2 + 4G^2}$ and $\tan\theta_p = 2G/\delta_B$. The superscript $(0)$ is used to denote that these are the unperturbed eigenstates and eigenvalues of $\hat{H}_{0}$. In the basis of $\hat{H_0}$ eigenstates, the full Hamiltonian is $\hat{H}'_{avg} = \hat{H}'_0 + \hat{V}'$ with 
\begin{equation}
 \hat{H}'_0  =  \begin{matrix}
|\chi^{(0)}_1\rangle\\
|\chi^{(0)}_2\rangle\\
|\chi^{(0)}_3\rangle
\end{matrix}
\begin{pmatrix}
(\delta_s + \omega)/2 & 0 & 0\\
0 & (\delta_s - \omega)/2 & 0\\
0 & 0 & \delta_-/2
\end{pmatrix} 
\end{equation}

and the perturbation 

\begin{equation}
 \hat{V}' =  
\begin{pmatrix}
0 & 0 & g c_p\\
0 & 0 & g s_p\\
g c_p & g s_p & 0
\end{pmatrix}
\end{equation}

 where $s_p = \sin(\theta_p/2)$ and $c_p = \cos(\theta_p/2)$.
 For a $2\times 2$ single spin system, PulsePol is resonant with the spin when the eigenvalues of $\hat{H}_{0}$ are degenerate, allowing $g$ to lift the degeneracy and couple the two eigenstates. Ignoring the blockade spin ($G = 0$), the resonance of the second spin occurs when $\delta_+ = \delta_-$, or $\delta_s = 0$. This is satisfied when the frequency of the PulsePol protocol $\omega_p$ is resonant with the spin's precession frequency, or

\begin{equation}
\omega_p = \omega_I^{(s)}
\end{equation} 

Using the definition of the protocol frequency $\omega_p = 6\pi/T$, the period of PulsePol, which is resonant with the nuclear spin, is $T_r = 6\pi/\omega_I^{(s)}$.

The same principle can be applied to the blockade spin system, where the degeneracy is lifted by the perturbation, $\hat{V}'$, when two eigenvalues of the $3\times 3$ Hamiltonian, $\hat{H}'_0$, are degenerate. Explicitly, $\hat{V}'$ weakly couples the state $|\chi_3^{(0)}\rangle$ to either $|\chi_1^{(0)}\rangle$ or $|\chi_2^{(0)}\rangle$. The eigenvalues of $\hat{H}'_0 $ are different depending on the sign of $\delta_-$. For example, if $\delta_- < 0$, then the eigenvalues $\epsilon_1^{(0)}$ and $\epsilon_3^{(0)}$ are never be degenerate, and $\hat{V}'$ is not strong enough to couple the corresponding states. However, there is a value of detuning in which $\epsilon_2^{(0)} = \epsilon_3^{(0)}$, giving an updated  condition for the weaker spin's resonance. For, $\delta_- > 0$ this is reversed and instead $\epsilon_1^{(0)} = \epsilon_3^{(0)}$. Initially, we will only consider systems where $\delta_- < 0$ and therefore the resonance satisfies $\delta_s - \sqrt{\delta_B^2 + 4G^2} = \delta_B - \delta_s$. Using the definition that $\delta_i = \omega_I^{(i)} - \omega_p$, then

\begin{equation}
\omega_p = \tilde{\omega}_I^{(s)} = \omega_I^{(s)} + \frac{G^2}{\delta_-}
\end{equation} 

which is shifted by $G^2/\delta_-$ from the original single spin resonant frequency $\omega_I^{(s)}$. Assuming that the coupling $g \ll G \ll \omega_L$, then the relative displacement of the resonant period is

\begin{equation}
\frac{\Delta T_r}{T_r} \simeq -\frac{G^2}{\omega_I^{(s)}(\omega_I^{(B)} - \omega_I^{(s)})}
\label{pol_shift}
\end{equation}

where the new resonance is at $T_r' \simeq  T_r + \Delta T_r$. The robustness of Eq.\eqref{pol_shift} is tested against both two spin simulations and experimental polarisation of C21 in Fig.3 of the main paper. 

\begin{figure}[ht!]\centering
		\includegraphics[width = 3.4in]{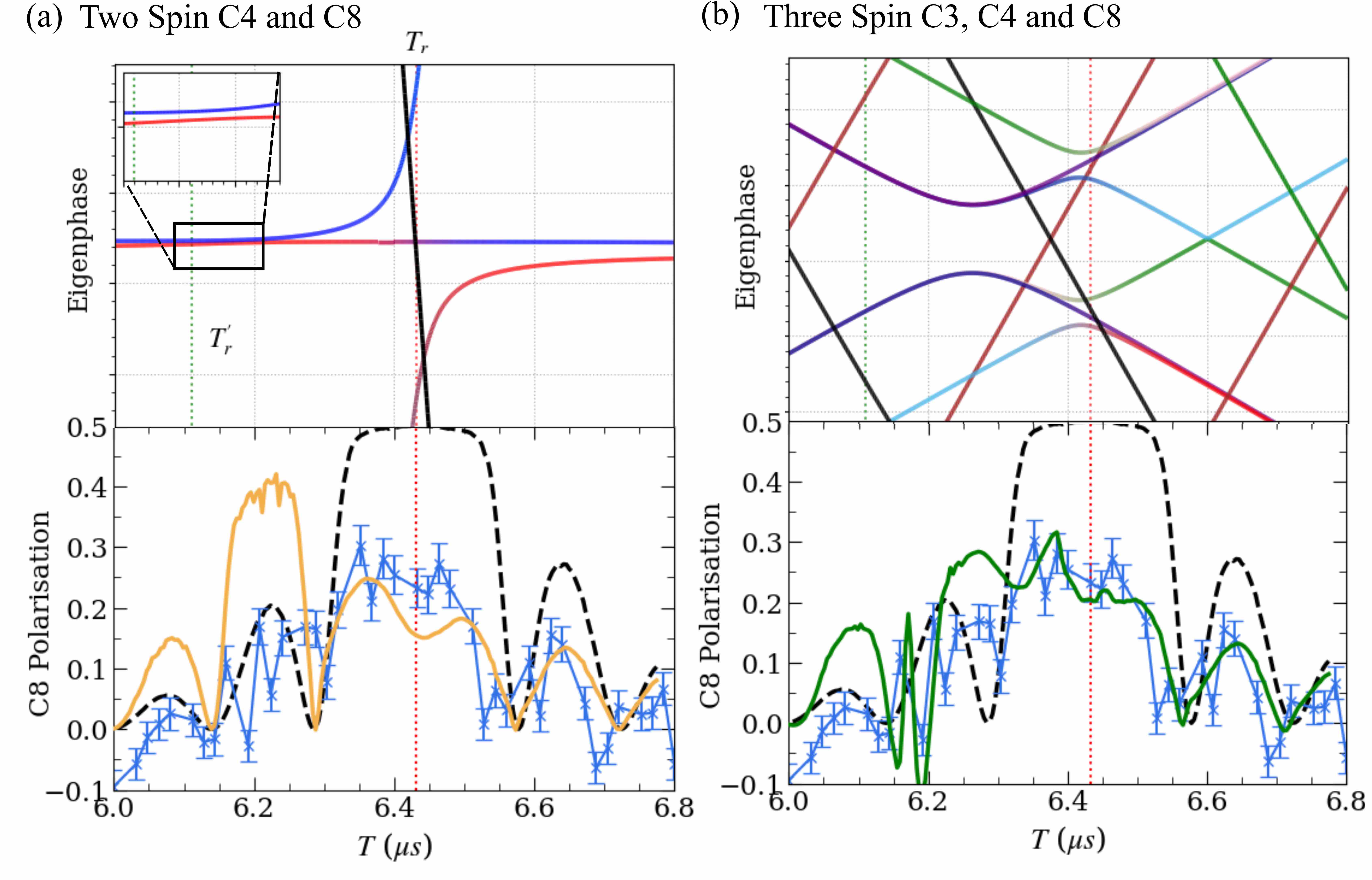}
	\caption{ Effect of two competing blockade spins: shows the polarisation of C8 at a higher harmonic ($k = 11$). (a) shows the Floquet phases and polarisation for a two spin system of C4 and C8. A displaced avoided crossing for C8 is present in the Floquet phases and a corresponding displaced polarisation peak from the single spin (black) to the two spin (orange). However, experimental data (blue) instead simply  has significantly lower polarisation  (b) shows Floquet phases and polarisation for a three spin system of C3, C4 and C8. The original displaced avoided crossing of C8 from (b) is now nearly degenerate with the avoided crossing of C3, which now acts as a blockade spin, pushing the crossing back towards its original position. This three spin simulation (green) demonstrates the flattening of the two spin peak and fits the experimental data better, showing the effect of higher order multi-spin effects. Simulations and data here are for $R = 1000$ repetitions of $N_p = 8$ PulsePol cycles.} 
	\label{Competing}
\end{figure}

Furthermore, degenerate perturbation theory can be used to find the Rabi-frequency of this resonance by solving the $2\times 2$ matrix in the subspace $\{|\chi_2^{(0)}\rangle,|\chi_3^{(0)}\rangle\}$, which is of the form $\hat{\tilde{H}}' = \epsilon\mathbb{I} + 2gs_p\hat{I}_x$ where $\epsilon = \epsilon_2^{(0)} = \epsilon_3^{(0)}$.  The corresponding Rabi frequency is found to be: 

\begin{equation}
\Omega_r \simeq 2g\sin\left(\frac{\theta_p}{2}\right)
\end{equation} 
This Rabi frequency is attenuated by a factor $\sin\theta/2$ relative to the one spin case. In fact, if $|\Delta T_r/T_r| \gg 1$ then the coupling between the two states will be completely suppressed and $\Omega_r = 0$.   However, for the typical scenarios studied here, the  displacement is less drastic  and  $\Omega_r \sim 2g$.
We note that in the $\delta_- \to 0$ limit, the shift tends to infinity.  In this scenario, the second small crossing vanishes and the dark mode behaviour is regained.

\subsection{Competing Blockade Spins}
While the focus here was primarily on the case of a single  spin (eg C3),  simultaneously and independently blocking multiple weaker-coupled spins,  we can consider the case where two blockade 
spins affect the same weaker coupled spin. This results in  a competing or combined blockade effect.  Such a scenario is identifiable in the experimental cluster here, when looking at higher harmonic resonances of PulsePol at larger $T$.

In previous sections,  a PulsePol resonance of harmonic $k = 3$ was studied. Consider now a higher harmonic, $k = 11$, at a much greater periodicity $T$. At higher harmonics, single spin resonances have a greater spacing, meaning that C8 and C4 are relatively close but no longer degenerate. Hence, the polarisation of C8 no longer saturates due to dark states,  and is displaced by $\Delta T_r \simeq -0.32 \,\mu\mathrm{s}$.
A simulation of this higher harmonic displacement of C8 due to C4 is shown in the first panel of Fig.\ref{Competing}. However, this is not seen in the experimental data, where there is significantly less polarisation than expected in this region of $T$.  In fact, by considering the initial blockade spin C3 in simulations, as is done in the second panel of Fig.\ref{Competing}, polarisation levels closer to the experimental data are obtained.

As in the Floquet states in Fig.\ref{Competing}, the expected displacement of the avoided crossing for C8 due to blockade spin C4 is nearly degenerate with the avoided crossing of C3, which was previously shown to act as a blockade spin for others with small $A_x$. Therefore, this shift closer to C3 causes a secondary blockade effect on C8, displacing the resonance back towards its original position and `smearing' the polarisation across $T$. This double-blockade effect on C8 is a three-spin effect, further highlighting the importance of considering many-body effects when performing polarisation of clusters.
\end{document}